# Evidence-based Prescriptive Analytics, CAUSAL Digital Twin and a Learning Estimation Algorithm

## Dr. PG Madhavan

CEO, Jin Innovation
Seattle, USA April 2021
pg@jininnovation.com

#Causality #Dynamics #DigitalTwin #Causaldigitaltwin #IoT #Causalgraph #Learning #Neuralnetwork

**Abstract**: *Evidence-based Prescriptive Analytics (EbPA) is necessary to determine optimal operational set-points that will improve business productivity. EbPA results from "what-if" analysis and counterfactual experimentation on CAUSAL Digital Twins (CDTs) that quantify cause-effect relationships in the DYNAMICS of a system of connected assets. We describe the basics of Causality and Causal Graphs and develop a Learning Causal Digital Twin (LCDT) solution; our algorithm uses a simple recurrent neural network with some innovative modifications incorporating Causal Graph simulation. Since LCDT is a "learning" digital twin where parameters are learned online in real-time with minimal pre-configuration, the work of deploying digital twins will be significantly simplified.*

*A proof-of-principle of LCDT was conducted using real vibration data from a system of bearings; results of causal factor estimation, what-if analysis study and counterfactual experiment are very encouraging.*

Digital twin is a software representation of a physical object, be it a machine, a system of connected assets or a whole city. In a typical embodiment of a digital twin, measurements via Internet of Things (IoT) enliven the software representation. CAUSAL digital twin (CDT) goes beyond that and captures the dynamics of interconnections among assets in terms of cause and effect relationships.

The real purpose of developing and deploying a digital twin is to understand the "parameters of the underlying system" so that we can *develop a CAUSAL understanding* of what the measurements are telling us about the system *("what is causing what?")*. You cannot easily do randomized controlled trials (the gold standard) with industrial machinery to find cause and effect! Digital twin (DT) is where cause-effect determination can happen. In my opinion, there is a lack of **prescriptive analytics** today which can tell a business what to do so that operations and production can improve – which requires causal relationships to be quantified beyond just correlations.

Clearly, the first round of "display" digital twins provided a nice window into IoT data via a dashboard, etc. "Simulation" digital twins have already proved their value in machine design where Static and Structural aspects are important – one case is the design of a product using CAD/CAM techniques where stresses and thermal distribution are displayed, useful in visualizing and improving the design of a machine.

However, to improve operational aspects, we have to focus on the DYNAMICS of a system. In the specific case of "machine dynamics", we are interested in the kinetics and kinematics of the machine and NOT the Statics or Structural aspects. *In this article, we develop a Causal Digital Twin (CDT) solution that captures the dynamics of connected systems.*

## Causality & IoT

Causal Digital Twin (CDT) provides us with a model on which we can perform "what-if" analysis which is the basis of "set point" optimization for operations, say in an industrial plant (or any other system of connected assets). This activity is called Prescriptive Analytics. *In short, Causal digital twin makes prescriptive analytics possible that can lead to improvements in operations and hence generate business value from Internet of Things (IoT) investments.*

All of today's Machine Learning (ML) is correlation-based. And we know that "correlation is NOT causation"! Most of the credit for the current "causality crusade" goes to Judea Pearl (2011 Turing award winner) who started his exposition of Causality Calculus from early 2000; Peter Spirtes' work in causality is equally impactful. Most of the Causality work has been in the field of social sciences – epidemiology, econometrics, genetics, market research, etc. This history has made it difficult for Engineers to access this vast knowledgebase and translate them for IoT use cases; *one of my main aims here is to simplify this process.*

In the rest of this article, I rely on the first two research publications below to formulate CDT. There is a large volume of research and I apologize in advance to the authors of other key articles that I may have missed. Third reference below discusses causality and root-cause analysis.

1. [Causal discovery and inference: concepts and recent methodological advances](#) (2016)
2. [Estimation of a Structural Vector Autoregression Model Using Non-Gaussianity](#) (2010)
3. [Causal Analytics in IIoT – AI That Knows What Causes What, and When](#) (2018)

A major source for everything related to causality is [Center for Causal Discovery.](#)

## Causality Basics

Let us start from the traditional Causality school's approach. At the risk of over-simplifying, here is a way an IoT engineer can come to grips with the basic tenets of Causality.

*The focus here is on the STRUCTURAL aspects and not sensor data time series generated by each asset! This is a key point to keep in mind – we will bring time dependencies later on.*

Consider 4 entities (assets), B1, B2, B3, B4, as shown in figure 1. The questions to ask are (1) Does all the links exist? (Some entities may not be connected); (2) If so, what is the direction of the link? And (3) What is the strength or weight of the links? The first 2 questions relate to "Causal Discovery" (of the structure) and the 3rd question relates to "Causal Estimation" (statistical/ signal processing methods when data are sensor-measured time series from each of the 4 assets).

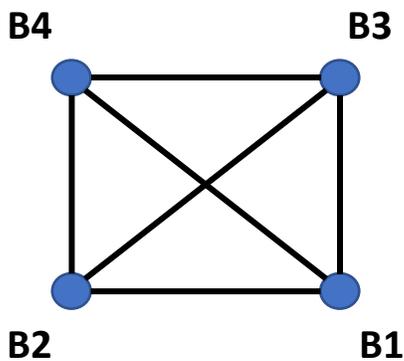

Figure 1. Fully connected graph

At the risk of oversimplifying, these are some of the main concepts in causal structure discovery.

- A causal model is *sufficient* if it does not contain unobserved common causes or latent variables.
- *Markov* assumption: For causally sufficient sets of variables, all variables are independent of their non-descendants in the causal graph conditional on their direct causes (parents in the causal graph).
- *Faithfulness*: Causal influence is not hidden by coincidental cancelations.
- In directed acyclic graphs (DAGs), one can use "d-separation" ("d" for directional) to identify pair-wise independent nodes which means that there is no link connecting the pair. d-separation is a mechanical procedure to answer the equivalent conditional independence question.

These factors have to be satisfied for Causal discovery and estimation to be valid – which is very challenging to prove in traditional applications in Social Sciences. For example, if the nodes of the graph are Health, # cigarettes smoked, Age and Obesity, one can imagine how difficult it will be to draw the correct directional links of the DAG and asserting Markov and Faithfulness assumptions!

## Causality Insights

Classification and regression and causal inference are different, according to Peter Spirtes who is a long-time leader of Causality at CMU; he presents the probabilistic reasoning for this in the 1st article cited earlier. There are important differences between the problem of predicting the value of a variable in an unmanipulated population from a sample (classification and regression) and the problem of predicting the post-manipulation value of a variable from a sample from an unmanipulated population (causality); the latter is called "counterfactual" analysis which reveals the difference.

- Causal factors in a DAG are pair-wise regression coefficients when Markov and Faithfulness conditions are satisfied and d-separation applied to the graph.

- The direction of the link (regress X on Y or Y on X) can be addressed by estimating the regression coefficient in each case, generating the residuals and applying statistical test for independence. If X truly causes Y, the residuals for the regression in this direction will be independent. Regression in reverse will generate residuals that are uncorrelated (by definition) but NOT independent.

Once given a DAG with the conditionally independent links between the nodes removed by d-separation and the direction of the link determined by checking the residuals, we have the Causal Graph as in figure 2. In practical use cases, we will have an outcome variable of interest; for example, in a market research study on a soap brand, the reported customer satisfaction or acceptable price may be the outcome variable and other nodes may be color, smell, size, etc. The market research team will pass around questionnaires to many people and collect these data which form the sample data at each node. Figure 2 is an example of a DAG with causal directions determined with B1 as outcome variable shown circled.

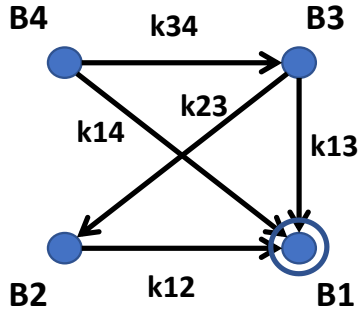

Figure 2. A causal graph with B1 as the outcome variable

From the market research study data collection example, it must be clear that TIME is not a variable in this causal analysis – this is called "structural" model or "instantaneous" causality. Consider the soap brand marketing study – measurements are from different consumers which may have been done at different times but the time-stamp has no bearing on the analysis. Such Structural Causality is the realm of "traditional" causality. In a general case though, the question arises of delayed or "lagged" measurements and their causal dependence.

In the IoT case, the factors influencing the outcome variable will have temporal dependencies in addition to the structural ones in general. Since IoT measurements are typically time series measured from multiple assets interconnected on a plant floor say, instantaneous and lagged causality are important. If the locations of each asset in the layout are significant, structural model is spatially distributed. *In the most general case, IoT data for DAG is a spatio-temporal multichannel time series.*

We are dealing with man-made systems in IoT. We will be working with a NASA Bearing Data case study later; a mechanic who has spent a lifetime working with bearings can tell you the key sources of vibration and how the sources are causally related. *As a significant simplification for this expositional study, we will use the approach that domain experts have provided us the corresponding DAG and link directions already during an initial knowledge-discovery phase*. Thus, our focus below will be on Causal Estimation.

# Real Data Modeling: NASA Bearing data

We will develop Causal DT in a real-life setting using the popular NASA Prognostics Data Repository's bearing dataset. The data is from a run-to-failure test setup of bearings installed on a shaft. The arrangement is shown in figure 3.

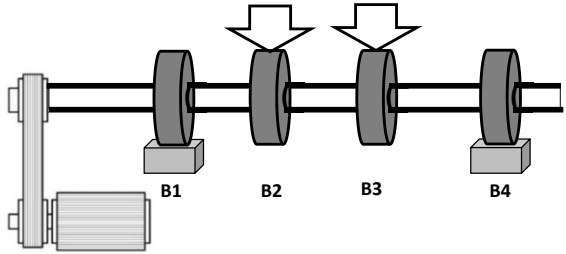

Figure 3. NASA Prognostic Data Repository Bearing data collection setup

There were 3 separate tests in the NASA dataset; we focus on Test 2. Data were collected from Feb 12, 2004 to Feb 19, 2004 and the tests were run continuously to failure in blocks of 10 minutes through the entire period. Bearing 1 will be the target of our study – its outer race failed on Feb 19.

From the structure of the setup and some knowledge of machine dynamics, bearing operation and vibration analysis, we can come up with the Causal Graph for what we will call "NASA bearing system".

# Recursive Estimation of Causal Parameters

Simulation results from an earlier recursive estimation study were reported here.

For the current study, we base our estimation method on Reference 2 in the initial section which addressed the estimation of instantaneous and lagged causal effects using Structural Vector Autoregressive (SVAR) models. There are many more significant papers in social sciences on this topic which IoT engineers can draw from.

General from of SVAR model:

$$\mathbf{y}_t = \mathbf{A}^0 \mathbf{y}_t + \sum_{m=1}^{M} \mathbf{A}^m \mathbf{y}_{t-m} + \mathbf{e}_t$$

All bolded quantities are vectors or matrices. $\mathbf{A}^0$ has zeros for its diagonal entries since self-causality does not exist. $\mathbf{y}_t$ are the measurements of the nodes of DAG. First term is Structural causality and the second term is Lagged causality.

We can rewrite SVAR model in a more general form in discrete time -

$\mathbf{y}[n] = \mathcal{F}^C \left[ k_{e,c}, \mathbf{y}[n] \right] + \mathcal{F}^D \left[ k_{e,c}^m, \mathbf{y}[n-m] \right]$ for m=1 to M  … (A)

$k_{e,c}, k_{e,c}^m = 0$ when e=c

$\mathcal{F}^C[.]$ – function of Current/ Structural samples of [.]

$\mathcal{F}^D[.]$ – function of Delayed/ Lagged samples of [.]

In this formulation shown in equation (A), the function, $\mathcal{F}$, can be linear or non-linear. The condition, $k_{e,c}, k_{e,c}^m = 0$ when e=c, means that a node has no "self-causality", i.e., a node does not cause any effect on itself at the same or at any lagged instant of time.

For estimation, separating equation (A) into 2 steps –

1. Find the mapping between $\mathbf{y}[n]$ & $k_{e,c}^{0\ to\ M}$

$$k_{e,c}^{0\,to\,M} = \mathcal{M}\{\mathbf{y}[n],\ g(\mathbf{y}[n-m])\}$$

$\mathcal{M}\{.\}$ – Neural Network map; $g(.)$ – function of Recurrent delayed **y**'s

2. Simulate causal graph

$$\hat{\mathbf{y}}[n] = \mathbf{A}^0\,\mathbf{y}[n] + \sum_{m=1}^{M} \mathbf{A}^m\,\mathbf{y}[n-m]$$

where **A**'s contain $k_{e,c}^{0\,to\,M}$ after NN convergence

Using a matrix version of a multilayer perceptron neural network (full derivation in the Appendix), we can learn all the $k_{e,c}^{0-M}$ simultaneously with some new modifications to the traditional multilayer perceptron. ***There are a few important features to notice in Figure 4.***

1. Neural network is an identity mapper, i.e., the input vector and the desired output vector are the same, **y**[n].
2. Input layer is augmented with a "context" layer (shown dotted). They contain delayed samples of some or all of the Hidden layer outputs. This recurrent neural network (RNN) was proposed by [Elman](). Our purpose here is to provide a mechanism to add a memory feature to the neural network.
3. *A significant departure from typical multilayer perceptron is the explicit addition of a Simulation layer* after the Output layer of the neural network. We will see the purpose of this variation when we discuss figure 5.
4. The usual error backpropagation algorithm will have to be modified to update the weights of Hidden and Output layer. This is accomplished by explicitly finding the error gradient across Simulation layer by perturbation analysis. [Complex-step derivative approximation]() method which has outstanding numerical properties are used here. The rest of backpropagation is traditional.

**Identity Mapping Recurrent Neural Network (of Elman type) over Simulation Layer**
*Error back-propagation over all layers algorithm*

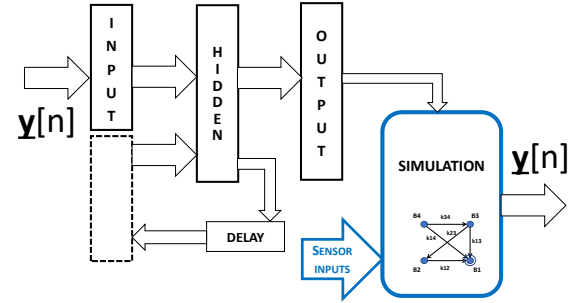

Figure 4. Structure of Identity-mapping recurrent neural network with simulation (IMRNNS)

The special features of our "Identity-mapping Recurrent neural network with Simulation"(IMRNNS) layer make the recursive estimation of all causal factors, $k_{e,c}^{0-M}$ possible. The algorithm is easier to follow in figure 5 where the essential operations alone are shown.

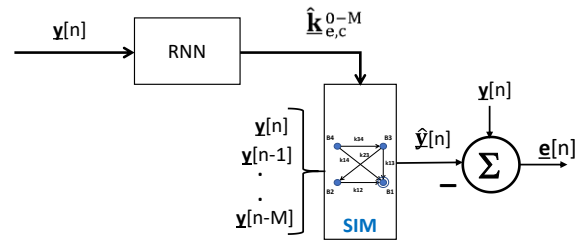

Figure 5. Essential operations of IMRNNS

To minimize the error, **e**[n]'s norm, $\hat{y}[n]$ has to approximate **y**[n] which requires RNN to provide the causal factor estimates, $\hat{\mathbf{k}}_{e,c}^{0-M}$, to the Simulation layer. Simulation layer in turn uses them to produce the estimate, $\hat{y}[n]$, that will minimize the error norm. At convergence, for n → large #, the best estimate of all $k_{e,c}^{0-M}$ is available as the output of RNN.

## Proof of Principle

For our purposes of demonstrating Causal Estimation for the NASA Bearing IoT use case, we will use the simplified DAG in figure 6 as well as assume that Markov and Faithfulness criteria for Causality have been met.

DAGs for all 4 bearings are shown in figure 6. Using IMRNNS, recursive estimation of all causal factors for all bearings is done simultaneously.

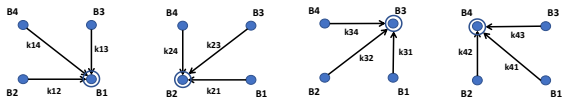

Figure 6. DAGs with Bearings 1, 2 3 & 4 as the Outcome Nodes of interest

Instead of using the entire time series from Feb 12 to Feb 19 when Bearing 1 failed (measured continuously over 24-hour periods), we use a block of 10 minutes from each significant date. The time series of Bearing 1 vibration measurement is shown in figure 7.

In figure 7, note that the data is NOT contiguous but 10-minute chunks from dates shown put together end to end. The causal factor estimation using IMRNNS was done for each day separately. From figure 7, it is obvious that Bearing 1 vibration is excessive on Feb 19 (bearing failed) and on Feb 12, we see the normal vibration pattern. As one gets closer from Feb 16, 17 and 18, one can see clearly that something is going wrong with Bearing 1 with Feb 16 data being slightly indicative of an impending fault.

*If the ONLY objective is to predict failure, simply thresholding the vibration amplitude may be enough of a solution* (but will not be very robust due to the presence of noise, etc.).

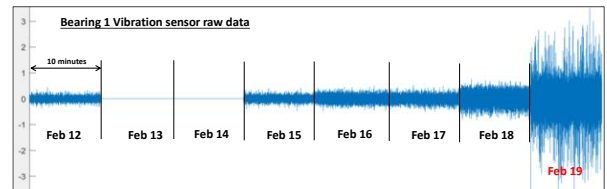

Figure 7. Time series of Bearing 1 vibration (NOT contiguous)

Using SVAR model and IMRNNS algorithm, we estimated $k_{e,c}^{0-M}$, the instantaneous and lagged causal factors for all 4 bearings simultaneously. We will call $k_{e,c}^{0-M}$ "coupling factors" to be more consistent with machine dynamics terminology.

In figure 8, we isolate the behavior of bearing, B1, since it is the one that failed. As can be seen, some of the couplings to B1 increased in magnitude on Feb 18 (and even on Feb 16). On Feb 19, the overall instability of the system when B1 failed seems to have broken up the coupling behavior (or it could be the nature of the time series that made the estimates very brittle).

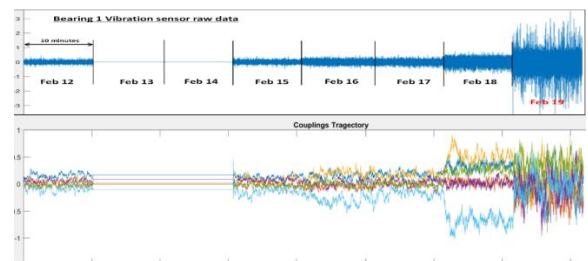

Figure 8. Bearing 1 time series (top) and all of its 6 couplings (bottom)

In figure 9, we shows all couplings for all bearings up to Feb 18. Significant couplings are called out as labels.

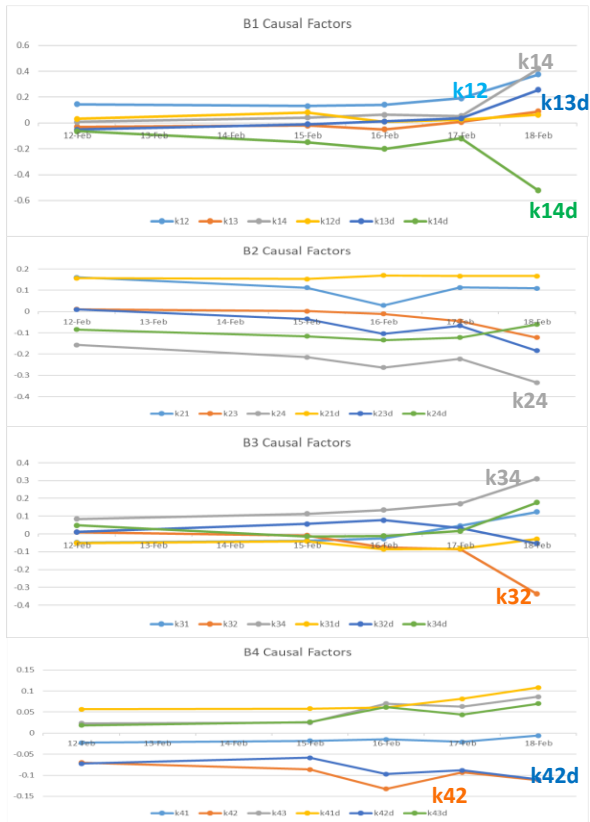

Figure 9. Causal coupling factors for all 4 bearings ("k": significant factors; "k-d": for Lag 1)

A key observation for Bearing B1 is that instantaneous couplings increase before failure; 2 lagged couplings show significant increases - figure 9, top panel. This can be seen much more clearly in figure 10 for bearing, B1, which failed.

Findings:

- Compared to using just the vibration amplitude, couplings show a systematic increase as Bearing 1 failure approached which opens up the possibility of a multi-factor failure prediction solution. How early a prediction could have been made is not visible in this study because only a 10-minute block of data per day was processed (Feb 17 data block chosen at random from 144 blocks on that day is suspect). If all continuous data were processed, it is very likely that the trajectories of $k_{e,c}^{0-M}$ will be smooth and thus be a robust basis for predicting failure point.

- Machine dynamics experts will take a special interest in delayed couplings; this is because the *flexing of the shaft can be the root cause of failure and the delay accounts for a possible back and forth feedback dynamic* between Bearing 1 and the other bearings (especially B4) on the same shaft. This learning can lead to better design of the bearing arrangement or stiffer material for the manufacture of the shaft.

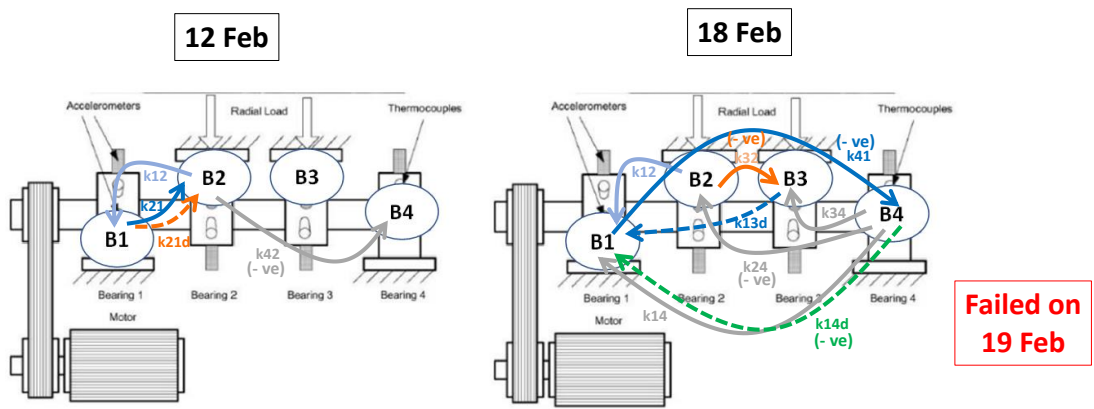

Figure 10. Visualizing couplings on a CAUSAL digital twin

We will remark that we also investigated simple factors such as variance ratios of pairwise bearing vibrations, etc.; no insightful information could be gathered.

In summary, it is very clear from figure 9 and 10 that couplings on Feb 18 are predictive of failure on Feb 19; it is noticeable that some of the couplings already show abnormal behavior as early as Feb 16. Figure 10 also makes a very informative "display" digital twin. Of course, these are not the primary reasons for Causal digital twin!

*In addition to visualizing the IoT data from the bearings in a new way (as in figure 10), the real value of a causal model is to perform "What-If" analysis and Counterfactual experiments*. We demonstrate the power of SVAR model in this regard in the next section.

## What-If analysis

SVAR digital twin provides a Causal Graph model with underlying parameters estimated from measured data. *With a "converged" model in hand, one can perform off-line experiments AFTER the data have been collected* – such as varying the parameters to assess the impact on Bearing 1. A simple study was undertaken to assess the "what-if" analysis capability of SVAR Causal digital twin.

Let us say that we are on Feb 12. We have all the vibration data on that day and the model in the last section. We varied 2 main delayed couplings, k13d and k14d, (shown in figure 11, top right) such that Bearing 1 data output of this simulation were similar to Bearing 1 data on Feb 18, the day before failure. The value of this "what-if" analysis is that if lagged k13 and k14 reach those levels on Feb 13, 14, . . ., we can conclude that Bearing 1 will likely fail the next day.

Clearly, one cannot compare Feb 18 Bearing 1 real data time series to the one simulated on Feb 12 – being realizations of random processes, they will not have the same waveforms. We use a more sophisticated method called Time-Frequency Distribution (TFD) which is a type of time-varying power spectral density estimate. Once the time axis is expanded, power variations over frequency can be visualized more clearly and some qualitative assessments can be made.

Figure 11 shows the results of What-If analysis described above. A very low-resolution TFD was estimated for the purposes of this experiment. As the legend explains, all diagrams here are TFDs of vibration data of Bearing 1. In figure 11 (A) and (C), the actual measured sensor data

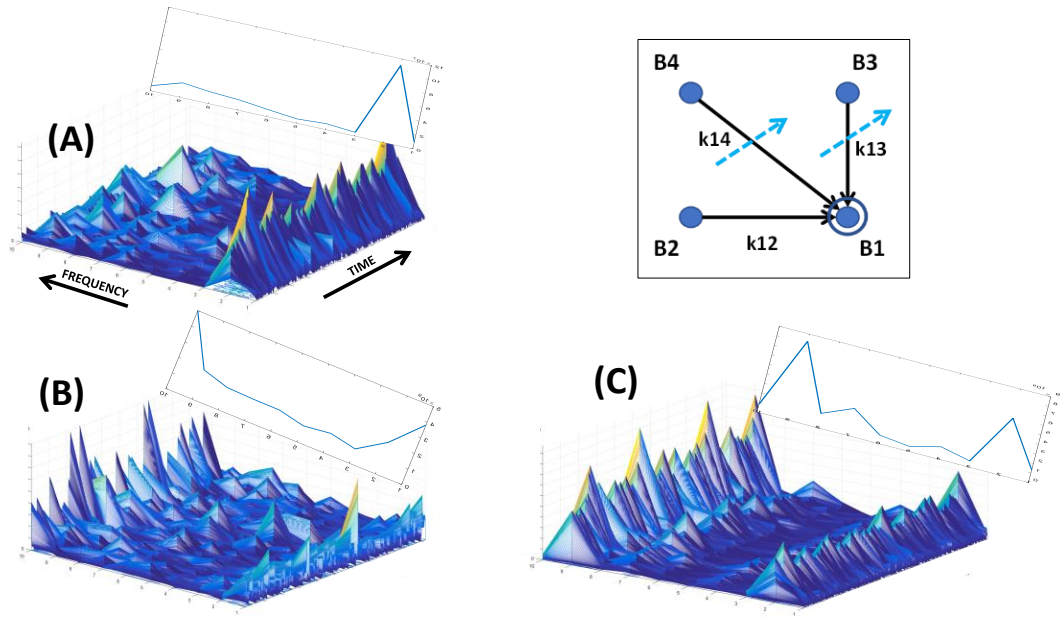

Figure 11. What-If analysis using SVAR causal digital twin
**(A)** Feb 12 Bearing 1 real data; **(B)** Simulated Feb 18 Bearing 1 data using Feb 12 Bearing 2, 3 & 4 real data as inputs with couplings varied; **(C)** Feb 18 Bearing 1 real data

TFDs from the NASA dataset are displayed. Comparing them, you can see that Feb 12 TFD in (A) has very little high-frequency content (frequency axis increases from right to left in all diagrams in figure 11). Whereas in (C), one can see a row of high-frequency peaks all across the 10-minute time axis.

This is made even more clear when you collapse all the time slices on to the frequency axis (thus yielding a traditional power spectrum estimate plot) which are shown floating on the right and top of each mesh plot. (C) has a clear high-frequency peak whereas (A) does not (note that frequency increases to the left).

(B) is the result of "what-if" analysis. What if we changed the coupling factors on Feb 12 data? Will the simulated Bearing 1 vibration look like the day before failure (Feb 18)? We see in figure 11(B) that for couplings from Feb 18, Feb 12$^{th}$ data simulate Bearing 1 data that "look like" Feb 18 data – (B) has more high-frequency peaks than (A) in the TFD and the power spectrum in (B) looks more similar to (C) than (A).

In essence, what we have done in this "what-if" analysis is to take Feb 12 data and see if the data will simulate a day before failure (Feb 18), which it does to some extent – *the value of this analysis is that we can simulate what may happen to the bearing system in the future*!

Another fascinating use is to run SVAR Causal digital twin in "fast forward" mode and explore what other effects it may exhibit in the future. In a more complex system than the relatively simple NASA bearing system, knowing the internal dynamics as specified by the Causal digital twin will have many more uses in assessing future state of the overall system and its individual components, remaining useful life (RUL) or simply predicting failure using causal graph parameters for a more robust (less false-positives and hence waste) method.

## Counterfactual experiment

The discriminating power of causal models is the ability to perform counterfactual experiments AFTER the data has been collected. To provide a demonstration, we assess the effect of the absence of Bearing 3 on Bearing 1. To eliminate Bearing 3, we set the instantaneous and lagged versions of coupling, k13, to zero (see figure 12, bottom left). The results for Bearing 1 data from Feb 18 (the day before failure) is shown in figure 12.

In figure 12, Bearing 1 TFD for Feb 18 and Feb 12 are obtained the same way as in Figure 11 (A) and (C). If on Feb 18, if Bearing 3 was not in the system, what will the vibration of Bearing 1 look like? The top right plot in figure 12 shows that the Bearing 1 vibration will be more similar to Feb 12$^{th}$ Bearing 1 data! Notice the lack of high-frequency peaks compared to actual Feb 18$^{th}$ data on the top left of figure 12. Clearly, Bearing 3 with a load (3000 lbs) on it has a role in Bearing 1's failure.

The practical use of this experiment to a machine dynamics expert is unclear but this counterfactual experiment is designed to show the power of Causal Graph models in the form of a Causal Digital Twin. *We can ask questions "after the fact" and seek answers to what could have happened* . . .

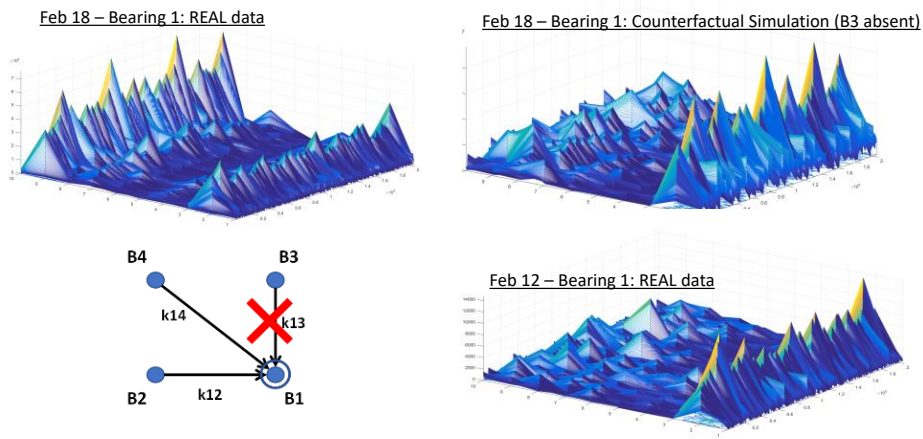

Figure 12. Counterfactual experiment using SVAR causal digital twin

## Significance of Causal digital twin

Structural Vector Auto-Regressive (SVAR) model and Identity-mapping Recurrent neural network with Simulation (IMRNNS) layer algorithm allow the estimation of causal parameters of a dynamical system from measurements. Causal Graphs that satisfy causality assumptions make such analyses possible.

- Causal graphs can be elicited from domain experts or estimated directly from measured data – "causal discovery" (there are startups providing this service – for example, https://www.inguo.io/). *In IoT use cases, domain-expert generated Causal Graph for causal discovery will be an adequate starting point*.
- SVAR & IMRNNS provide a simple methodology for estimating instantaneous and lagged causal link strengths (or couplings).
- In the real data tests with NASA bearing data, we showed that SVAR Causal digital twin works in real-life cases and is able to unearth the physical system's underlying unobservable parameters by learning them over time.
- *Causal Digital Twin offers the unique ability to perform "what-if" and counterfactual experiments on already-collected data to develop insights into the operation of connected assets*.
- One of the most desirable features of SVAR Causal digital twin is that there is no hand-tuning of the digital twin. All steps are LEARNED iteratively as each data point arrives. Therefore, our Causal digital twin is a true "real-time" digital twin. *Once a causal graph structure is chosen, all relevant parameters are learned "on the fly", there by instantiating each physical system automatically*.

## Looking Ahead

*We present this work as a foundational expository application (perhaps the first one) of Causality theory to the DYNAMICS of connected system of assets*. There is a lot more to be borrowed from Causality theory – the references in the first section are great starting points. When it comes to estimation, SVAR model just scratches the surface.

Signal processing and Machine Learning experts have an extensive arsenal of recursive (real-time) methods based on conditional expectation estimation in a Kalman filter framework and other methods refined over the last 30 or so years. Interesting extension to nonlinear causal factor estimation in biomedical domain has already happened and I foresee an explosion of Causal digital twins suitable for various asset dynamics IoT use cases.

*The performance of "what-if" and counterfactual experiments using causal digital twins are the sources of actionable information for evidence-based prescriptive analytics which can be a reliable basis for business decisions to improve operations and increase production.*

Causal digital twin (CDT) solution presented here focuses on a system of connected assets and its dynamics; this causality focus of ours may be unique in IoT asset dynamics applications so far. Addressing one asset at a time has its uses – in predicting that asset's potential failure and better structural design. However, when connected assets are treated as a system as used in a production environment, the results of ***Causal Digital Twin that learns in real-time become directly useful in enhancing business outcomes such as increased production volume, better quality and reduced waste – all contributing to increased gross margin for the business***.

**About the author:**

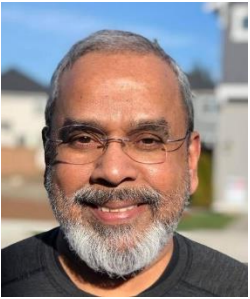

Dr. PG Madhavan launched his first IoT product at Rockwell Automation back in 2000 for predictive maintenance, an end-to-end solution including a display digital twin. Since then, he has been involved in the development of IoT technologies such as fault detection in jet engines at GE Aviation and causal digital twins to improve operational outcomes. His collected works in IoT is being published as a book, "Data Science for IoT Engineers" in June 2021. Rest of his career has been in industry spanning more major corporations (Microsoft, Lucent Bell Labs and NEC) and four startups (2 of which he founded and led as CEO). https://www.linkedin.com/in/pgmad/

APPENDIX:

**"Identity-mapping Recurrent neural network with Simulation"(IMRNNS) layer Algorithm Derivation:**

We derive a matrix version of multilayer perceptron neural network. "$\Psi$" is the Causal Graph simulation block. Elman's recurrence is not explicitly called out in this structure but it is a trivial extension via augmenting the Input Layer with additional "context" layer nodes with the effect that the number of input layer nodes, M, increases.

All variables are vectors or matrices in the derivation below.

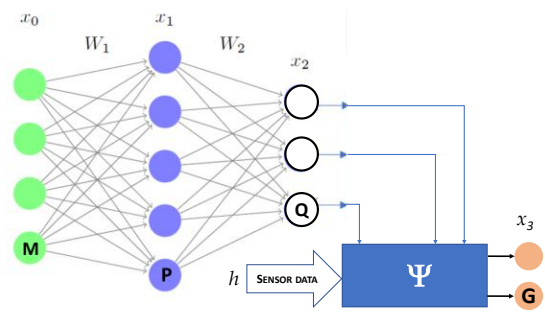

Neural Network: M – Input nodes; P – Hidden nodes; Q – Output ($x_2$) nodes; G – SIM Outputs ($x_3$).

$X_0$ – external inputs; Sensor Data, "h" – external measurements for graph simulation.

**NOTE:** *Output node $X_2$ values = Causal Factors (or "couplings")*

*For ease of implementation, vector-matrix dimensions are listed within brackets.*

Forward Pass:
$W_1$ – (PxM), $X_0$ – (Mx1) ; $X_1 = \rho [W_1 X_0]$ ; - (Px1)
$\rho [W_1 * X_0] = \frac{2}{1+exp(-W_1*X_0)} - 1.0$

$W_2$ – (QxP), $X_1$ – (Px1); $X_2 = W_2 X_1$ ; - (Qx1)

Function, $\Psi[\,.\,]$ takes a Qx1 vector and maps to a (Gx1) vector
$X_3 = \Psi[X_2] = \Psi[X_2, h]$ ; - (Gx1)

True output = t, (Gx1) vector
Error, $e = (t - X_3)$ ; (Gx1)
Error norm, $E = (e^T e)$ ; scalar

$\frac{\partial E}{\partial X_3} = 2e(-1) = -2e$ , - (Gx1)

$\frac{\partial X_3}{\partial X_2} = \Delta\{\Psi[X_2]\} = \Delta\{\Psi\} = $
$\left[ \sum_{i=1}^{G} \frac{\Delta X_{3i}}{X_{2j_{\neq 1}} \Delta X_{21}} \sum_{i=1}^{G} \frac{\Delta X_{3i}}{X_{2j_{\neq 2}} \Delta X_{22}} \cdots \sum_{i=1}^{G} \frac{\Delta X_{3i}}{X_{2j_{\neq Q}} \Delta X_{2Q}} \right]$
- (1xQ)

$\frac{\partial E}{\partial X_2} = -2e \otimes \Delta\{\Psi\}$
$= \left[ \begin{array}{c} \sum_{i=1}^{G} \frac{\Delta X_{3i}\, e_i}{X_{2j_{\neq 1}} \Delta X_{21}} \sum_{i=1}^{G} \frac{\Delta X_{3i}\, e_i}{X_{2j_{\neq 2}} \Delta X_{22}} \cdots \\ \cdots \sum_{i=1}^{G} \frac{\Delta X_{3i}\, e_i}{X_{2j_{\neq Q}} \Delta X_{2Q}} \end{array} \right]^T$, - (Qx1)

$\frac{\partial X_2}{\partial W_2} = X_1$ , (Px1) | $\therefore \frac{\partial E}{\partial W_2} = -2e\, \Delta\{\Psi\}\, X_1^T$
- (Qx1) x (1xP) = (Q x P)

$\frac{\partial X_2}{\partial X_1} = W_2$ , (QxP) | $\frac{\partial X_1}{\partial (W_1 * X_0)} = \Delta\rho$ , (Px1);
$\Delta\rho = 2 * X_1 \otimes (1 - X_1)$

$\frac{\partial (W_1 * X_0)}{\partial W_1} = X_0$ , (Mx1)

$\therefore \frac{\partial E}{\partial W_1} = -2 \{\Delta\rho \otimes W_2\} [e \otimes \Delta\{\Psi\}]\, X_0^T$ ,
- (PxQ) x (Qx1) x (1xM) = (PxM)
  {$\Delta\rho \otimes W_2$} - (PxQ); Column-wise Op; Multiply each column of $W_2$ by each element of $\Delta\rho$

Weight Updates:
$\Delta w_2 = -½\,\eta * \frac{\partial E}{\partial W_2} = \eta * [e \otimes \Delta\{\Psi\}] * X_1^T;$
$\Delta w_2$ – (QxP)

$\Delta w_1 = -½\,\eta * \frac{\partial E}{\partial W_1} = \eta * \{\Delta\rho \otimes W_2\} * [e \otimes \Delta\{\Psi\}] * X_0^T$
$\Delta w_1$ – (PxM)

$\Delta\{\Psi\}$ is obtained by perturbation analysis using [Complex-step derivative approximation](#) method.